\documentclass[%
reprint,twocolumn,
superscriptaddress,
amsmath,amssymb,
floatfix,
]{revtex4-1}

\usepackage{float}
\usepackage{graphicx}
\usepackage{dcolumn}
\usepackage{bm}
\usepackage[colorlinks,linkcolor=blue,anchorcolor=blue,citecolor=blue,urlcolor=blue]{hyperref}
\usepackage{longtable}

\begin{document}

\title{Robust optical-levitation-based metrology of nanoparticle's position and mass}
\author{Yu Zheng}
\affiliation{CAS Key Lab of Quantum Information, University of Science and Technology of China, Hefei 230026, China}
\affiliation{CAS Center For Excellence in Quantum Information and Quantum Physics, University of Science and Technology of China, Hefei 230026, China}
\author{Lei-Ming Zhou}
\affiliation{Department of Electrical and Computer Engineering, National University of Singapore, 4 Engineering Drive 3, Singapore, 117583, Singapore}
\author{Yang Dong}
\affiliation{CAS Key Lab of Quantum Information, University of Science and Technology of China, Hefei 230026, China}
\affiliation{CAS Center For Excellence in Quantum Information and Quantum Physics, University of Science and Technology of China, Hefei 230026, China}
\author{Cheng-Wei Qiu}
\affiliation{Department of Electrical and Computer Engineering, National University of Singapore, 4 Engineering Drive 3, Singapore, 117583, Singapore}
\author{Xiang-Dong Chen}
\author{Guang-Can Guo}
\author{Fang-Wen Sun}
\email{fwsun@ustc.edu.cn}
\affiliation{CAS Key Lab of Quantum Information, University of Science and Technology of China, Hefei 230026, China}
\affiliation{CAS Center For Excellence in Quantum Information and Quantum Physics, University of Science and Technology of China, Hefei 230026, China}

\date{\today }
\begin{abstract}
Light has shown up an incredibe capability in precision measurement based on opto-mechanic interaction in high vacuum by isolating environment noises. However, there are still obstructions, such as displacement and mass estimation error, highly hampering the improvement of absolute accuracy at the nanoscale. Here, we present a nonlinearity based metrology to precisely measure the position and mass of a nanoparticle with optical levitation under $10^{-5}$ mbar, 6-order of magnitude lower than the electrostatic-force and stochastic-force-based counterparts.
By precisely controlling the amplitude of the levitated nanoparticle at the nonlinear regime, we realized a feasible sub-picometer-level position measurement with an uncertainty of $1.0\%$ without the prior information of mass, which can be further applied to weigh the femtogram-level mass with an uncertainty of $2.2\%$. It will also pave the way to construct a well-calibrated opto-mechanic platform in high vacuum for high sensitivity and accuracy measurement in force and acceleration at the nanoscale and the study in quantum superposition at the mesoscopic scale.
\end{abstract}
\maketitle
Light has been the most powerful tool for precision metrologies in time, frequency, and distance \cite{Brewer2019time,Udem2002frequency,Yuan2019ruler}. Based on opto-mechanic interaction in high vacuum, the gravitational wave has also been successfully detected \cite{LIGO2017}. Recently, the compact optical levitation in vacuum, which joins the fields of optomechanics \cite{Aspelmeyer2014cavity} and optical trapping \cite{Ashkin1980radiation,Ashkin1986tweezer,Gao2017qiuchengwei}, is being put into the spotlight of high precision metrology for modern science, especially at the nanoscale. It extends the optical precision metrology in force
\cite%
{Ranjit2016force,Hempston2017force,Hebestreit2018freefall,Blakemore2019force}%
, mass \cite{Blakemore2019mass,ricci2019mass}, charge \cite{Moore2014charge}
and acceleration \cite{Monteiro2017acc}. Such a system has also been considered as a promising platform for the investigation of quantum superposition at the mesoscopic scale \cite{romeroisart2010quantum,Romero2011quantum,Bateman2014nearfield}, which may further enhance the performance of precision metrology due to the nature of quantum superposition and entanglement \cite{Giovannetti2011}.

\begin{figure}[t]
\includegraphics[width=0.45\textwidth]{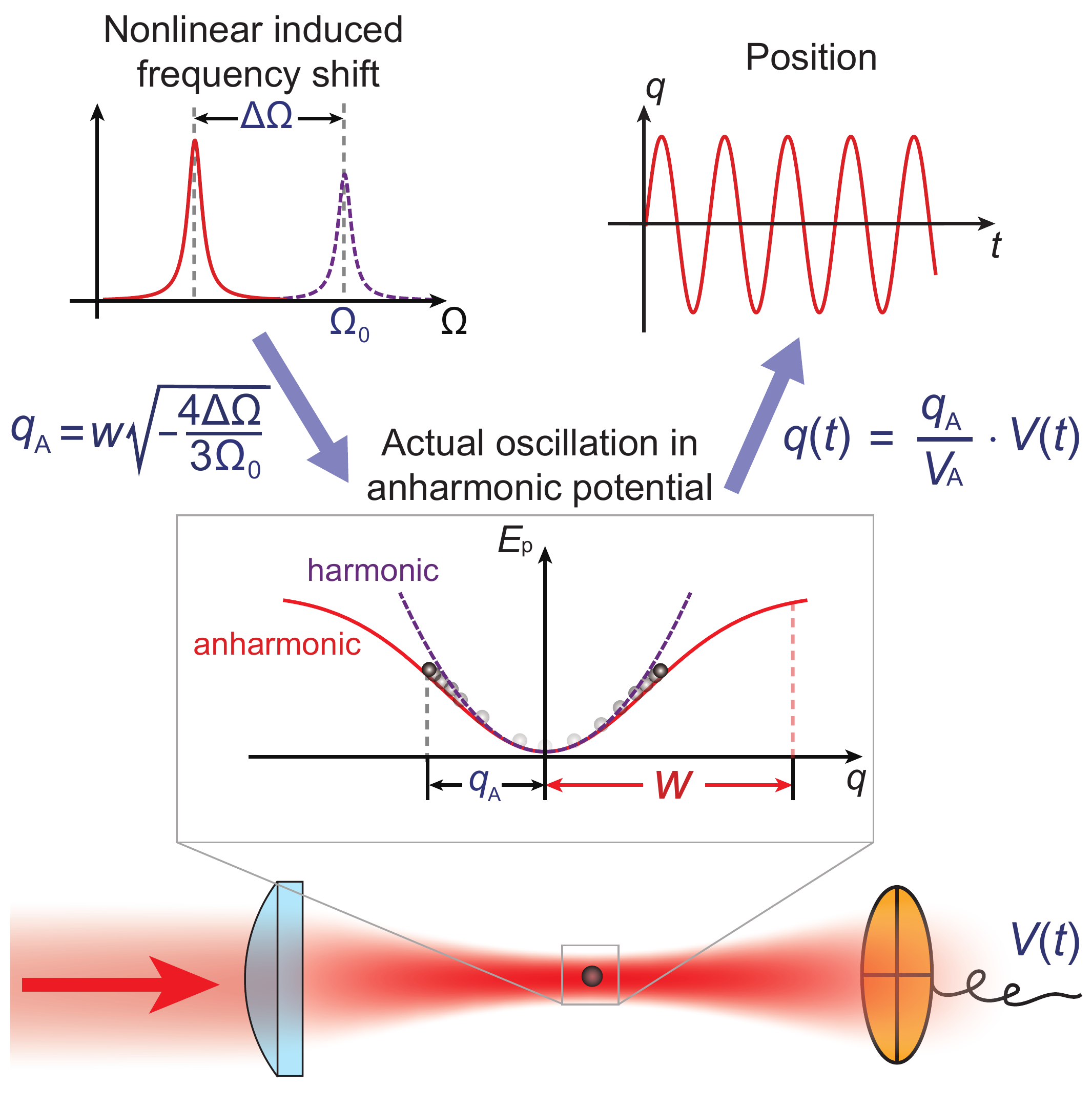}
\caption{Scheme of position measurement without the information
of particle mass. The measured nonlinearity induced relative natural
frequency shift ${\Delta\Omega}/{\Omega_{0}}$, which can be obtained from the spectrum analysis of the voltage signal of detector $V(t)$, is related to the actual oscillation amplitude $q_{A}$ as $q_{A}=w\sqrt{-4\Delta\Omega/(3{\Omega_{0}})}$.
With the respective measured voltage signal amplitude $V_{A}$, the
calibrated factor $c=q_{A}/V_{A}$ is got with only the information
about $V(t)$. Then the position signal can be obtained as $q(t)=c\cdot V(t)$. }
\label{fig:1}
\end{figure}

It had been regarded that the optical levitation has a high precision in position detection with state-of-the-art techniques, which required a priori knowledge of the particle mass \cite{hebestreit2018} or the assistance of stochastic \cite{berg2004power,Hauer2013Calib,hebestreit2018} or extra electrostatic \cite{hebestreit2018,ricci2019mass} forces at a moderate vacuum to calibrate the experimental output. However, in the experiment, the mass of the nanoparticle and stochastic and electrostatic forces can be hardly estimated or measured precisely at the nanoscale \cite{millen2014temp,Frangeskou2016diamond,Hebestreit2018temp},
which severely reduces the accuracy of the position and other metrologies with optical levitation system.

Here, we 
present an all-optical metrology for position and mass measurement with optical levitation in high vacuum. The geometry shape of the optical potential is regarded as a calibration gauge. The deviation from quadratic shape potential will induce a nonlinear natural frequency shift \cite{gieseler2013nonlinear}, which acts as the ruler mark to precisely read out the position and the motion of the nanoparticle without a priori knowledge of mass. Furthermore, we are able to weigh the mass and measure the density of the optically levitated nanoparticle. This high accuracy position and mass measurement can help to construct a well-calibrated opto-mechanic platform for high sensitivity and accuracy measurement in force and acceleration.
\begin{figure}[t]
\includegraphics[width=0.45\textwidth]{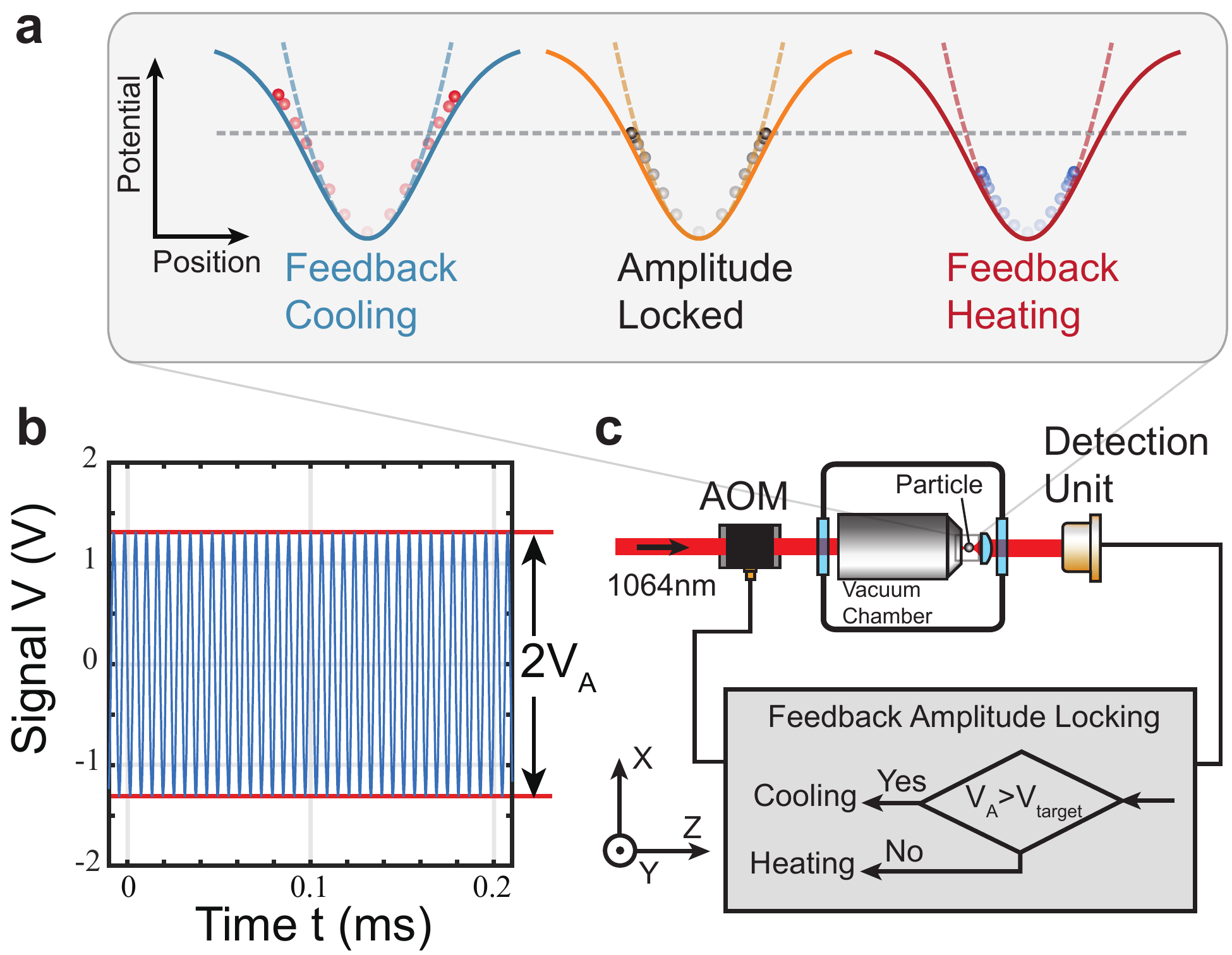}
\caption{(a) Schematics of the parametric feedback locking. When the measured amplitude ($V_A$) of the trapped particle is larger (smaller) than the target amplitude ($V_{target}$), modulations will be
applied with feedback cooling (heating) to reduce (increase) the amplitude.
The solid curves represent the nonlinear Gaussian potential which deviate
from harmonic potential (dashed curves). (b) Part of the recorded signal time
trace from an amplitude locked particle at a pressure of $10^{-5}$ mbar. (c)
Simplified experimental setup. A $1064$ nm laser was focused by an objective
which was mounted in vacuum chamber to form the optical potential for
particle trapping. The forward scattering light from optical trap was
collected and sent to c.m. motion detection unit. The detector signal was processed to generate
the feedback amplitude locking by modulating the trapping laser intensity
with an acousto-optical modulator (AOM).}
\label{fig:2}
\end{figure}

An optically levitated sensor realizes its functions by analyzing
the motions of the trapped particle. The optical potential is harmonic
and the oscillation has a fixed natural frequency $\Omega_{0}$ (shown
as the purple dashed potential shape in Fig. \ref{fig:1}) when the particle is
trapped very near the equilibrium point. However, in an optical trap
built by a laser beam in Gaussian mode \cite%
{gieseler2013nonlinear,novotny2012principles}, the optical potential
will become anharmonic when the oscillator has a large amplitude
and can move far from the equilibrium point (shown as the red solid
potential shape in Fig. \ref{fig:1}). This anharmonic potential features a
Duffing nonlinearity \cite%
{gieseler2013nonlinear}. Ignoring the interaction between different
spatial degrees of freedom, the motion of the trapped particle with mass $m$ in
this nonlinear regime can then be described as
\begin{equation}
\ddot{q}+\Gamma_{0}\dot{q}+\Omega_{0}^{2}\left( q+\xi q^{3}\right)=F_{therm}/m%
\text{,}  \label{motion}
\end{equation}
in one degree of freedom, where $q(t)$ is the position of the trapped
particle. $\Gamma_{0}$ is the damping rate induced by air. $\Omega_{0}$ is
the natural angular frequency when the oscillation is in the linear
regime. $F_{therm}$ is the Brownian stochastic force. $\xi$ describes the nonlinear coefficient of the
trap. For a Gaussian distribution optical potential, $\xi=-{2}/{w^{2}}$,
where $w$ is the $1/\mathrm{e}^{2}$ beam intensity radius. The position
signal $q(t)=c\cdot V(t)$ is expected to be experimentally measured
by a photodetector output voltage $V(t)$ and a calibration constant
$c$.

It is noticed that we can get the calibration constant $c$ with
only the voltage signal $V(t)$. Due to the nonlinearity, the
oscillator has got a natural frequency shift $\Delta\Omega=-\frac{3}{4}{w}^{2}q_{A}^{2}\Omega_{0}$
according to the theoretical solution of Eq. (\ref{motion}), when the amplitude
of the oscillator is $q_{A}$. As both $\Omega_{0}$ and $\Delta\Omega$
can be obtained from the spectrum analysis of $V(t)$, we can get $q_{A}$
without other information. Along with the respective measured voltage
signal amplitude $V_{A}$ , the calibration constant $c=q_{A}/V_{A}$
is determined. In more details, in experimental implementation to
obtain the calibration constant $c$ with high accuracy, we can fit the dependence of
the relative natural frequency shift on the voltage amplitude
\begin{equation}
\frac{\Delta \Omega}{\Omega_{0}}=\alpha V_A^{2}\text{,}  \label{alpha}
\end{equation}%
with a nonlinear coefficient $\alpha$. Then the calibration constant
$c=\sqrt{-\frac{4}{3}\alpha w^{2}}$, which can be figured out from
$\alpha$ and $w$.

In order to measure $\alpha $ based on Eq. (\ref{alpha}), we need to measure the frequency shift under different
voltage amplitudes. Although some existing methods can be employed, such as
picking out divided trajectory parts with different amplitudes of a
thermally driven particle \cite{gieseler2013nonlinear}, double-frequency
parametric driving \cite%
{gieseler2013nonlinear,Gieseler2014nonlinear,ricci2017bistable}, and optical
tweezer phonon laser \cite{Pettit2019phonon}, the imprecise amplitude
control severely limits the performance and reliability of such methods for
calibration.

\begin{figure*}[t]
\includegraphics[width=0.95\textwidth]{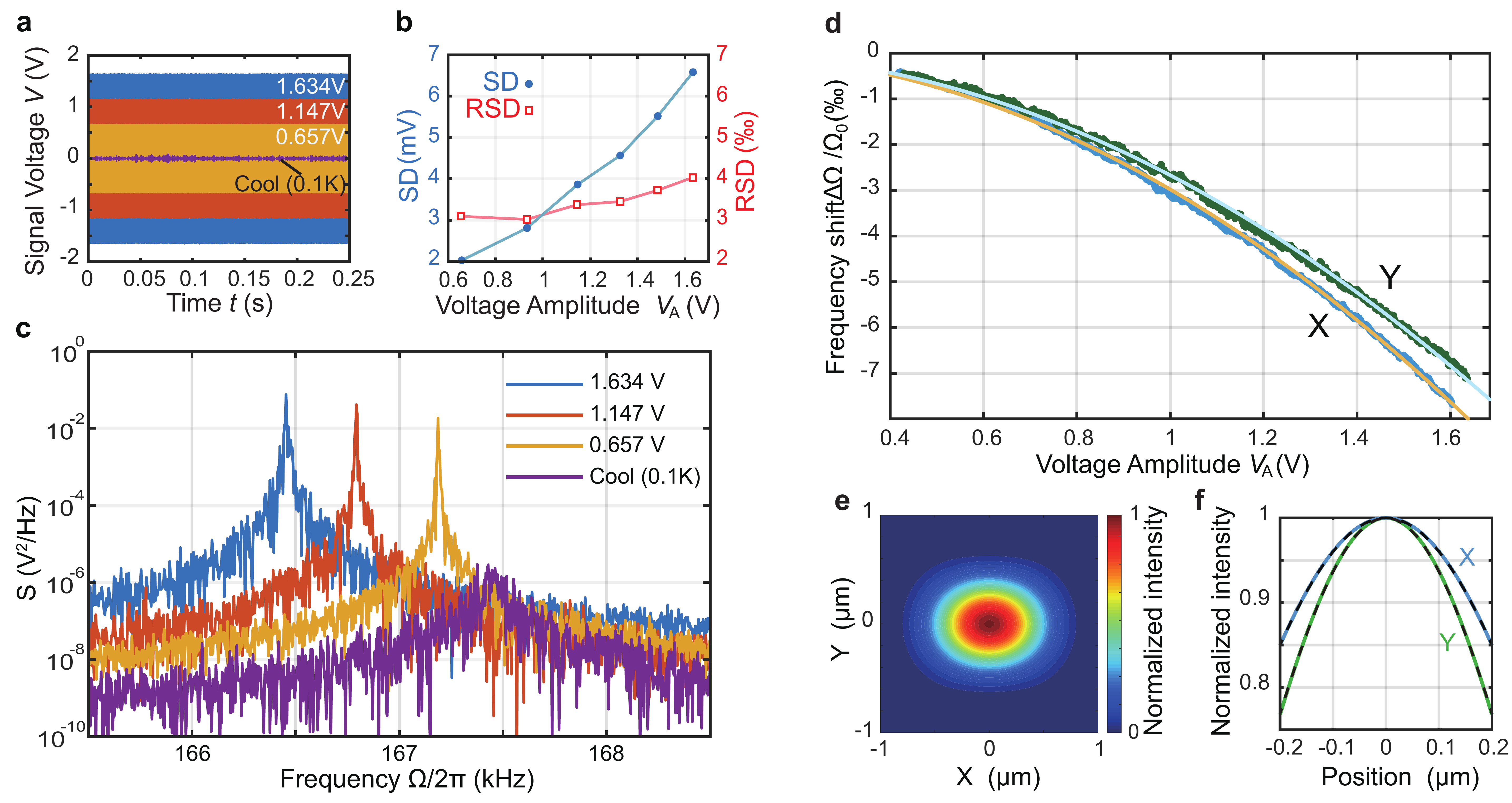}
\caption{(a) Measured signal voltage time traces of a trapped nanoparticle under
different locked amplitudes and feedback cooling along the Y axis at a
pressure of $10^{-5} $ mbar. (b) Dependence of the absolute (SD) and relative (RSD)
standard deviations of the voltage amplitude $V_A$. (c) Power spectral
densities of the voltage signals shown in (a). (d) Measured relative natural frequency shifts along X axis (blue) and Y (green) axes as a function of
voltage amplitude. The solid lines are fitted with Eq. (%
\protect\ref{alpha}). (e) The intensity distribution simulated with Debye integral in
the X-Y plane. (f) The simulated intensity distributions along the X-axis
(blue line) and Y-axis (green line). The dashed lines are fitted with Gaussian
function.}
\label{fig:3}
\end{figure*}
Here, we introduce a parametric feedback amplitude locking modulation to
precisely control the oscillation amplitude, which corresponds to the center of
mass (c.m.) motion temperature. As shown in Fig. \ref{fig:2}(a), when the
amplitude is higher or lower than a target amplitude, a parametric feedback
cooling \cite{Gieseler2012Cooling,Vovrosh2017Cooling} or heating \cite%
{zheng2019cooling,Vovrosh2017Cooling} process is applied to drive the
amplitude to the target. In detail, assuming the short term time trace of the
oscillation's voltage signal shown in Fig. \ref{fig:2}(b) can be described as $u_{\sin }(t)=V_A\sin \left[
\Omega _{0}t+\varphi \right] $, where $\varphi $ is the phase of the
oscillation. And a $\mathrm{\pi }/2$-phase-shifted signal is $u_{\cos
}(t)=V_A\cos \left[ \Omega _{0}t+\varphi \right] $. The applied trapping laser
intensity for feedback amplitude locking modulation is
\begin{equation}
I(t)=I_{0}\left\{ 1+A(t)\times 0.5\eta _{\text{m}}\times \mathrm{sign}\left[
u_{\sin }(t)u_{\cos }(t)\right] \right\} \text{,}  \label{modulation}
\end{equation}%
where $I_{0}$ is the laser intensity without modulation and $\eta _{\text{m}%
} $ is the modulation depth \cite{zheng2019cooling}. $A(t)=\mathrm{sign}( V_A^{2} -V_{\text{target}}^{2})$ is the amplitude
criterion.
When $V_A$ is larger (smaller) than the target voltage amplitude $V_{\text{%
target}}$, $A$ is positive (negative) and Eq. (\ref{modulation}) will be a
feedback cooling (heating) modulation to decrease (increase) the amplitude
of the levitated oscillator \cite{zheng2019cooling}.

In the experiment, as shown in Fig. \ref{fig:2}(c), an objective ($\mathrm{NA}=0.9$%
) was mounted in the vacuum chamber to create an optical potential by
focusing the Gaussian trapping laser. To avoid the interaction between motional degrees
of freedom when more than two axes of oscillation reach the nonlinear
regime \cite{gieseler2013nonlinear,Gieseler2014nonlinear}, only one axis motion will be amplitude-locked at a time, and the motions of the other two
axes will be cooled to sub-Kelvin \cite{zheng2019cooling}. The modulation depth $ \eta _{\text{m}} $ was set to be 0.5\% during the experiment.

Figure \ref{fig:3}(a) shows the measured signal voltage time traces of a trapped particle under
different locked amplitudes and feedback cooling along the Y axis at a
pressure of $10^{-5} $ mbar. The relative standard
deviation of the voltage amplitude for such an amplitude-locked oscillator was lower than $0.5\%$, as shown in Fig. \ref{fig:3}(b). And the power
spectral densities (PSD) with different locked amplitudes confirmed that the
natural frequency of the oscillator in an anharmonic optical potential
decreased with increasing amplitude, as shown in Fig. \ref{fig:3}(c).

Based on the feedback amplitude locking  technique at the nonlinear regime, the oscillator frequency was measured while the amplitude
locking target was swept from $0.4$ $\mathrm{V}$ to $1.6$ \textrm{V}. And the measurement results were averaged from tens of
sweep cycles to eliminate the noise and low-frequency drift due to the system
instability. This measurement procedure was applied to both $X$ and $Y$ axes to verify the reliability of the calibration process.
The dependence of relative
frequency shift on voltage amplitude is shown in Fig. \ref{fig:3}(d). By
fitting it with Eq. (\ref{alpha}), we can get that the nonlinear coefficients of $%
X $-axis and $Y$-axis are $\alpha _{\mathrm{X}}=-0.002978\text{ }\mathrm{V}%
^{-2} $ and $\alpha _{\mathrm{Y}}=-0.002669\text{ }\mathrm{V}^{-2}$,
respectively.

To complete the high precision calibration, the intensity distribution in the $X-Y$ plane was simulated with Debye integral \cite%
{gieseler2013nonlinear,novotny2012principles}, as shown in Figs. \ref{fig:3}%
(e) and (f). By fitting the intensity along $X$ and $Y$ axes with Gaussian
function, we get the $1/\mathrm{e}^{2}$ intensity radius $w_{\mathrm{X}%
}=703\pm 7\text{ }\mathrm{nm}$ and $w_{\mathrm{Y}}=551\pm 22\text{ }\mathrm{%
nm}$, respectively. The errors come from the uncertainty of optical
component parameters (see Supplementary Information for details \cite{SI}). With $\alpha _{\mathrm{X}}$, $\alpha _{\mathrm{Y}%
}$, $w_{\mathrm{X}}$, and $w_{\mathrm{Y}}$, we
finally extracted the calibration constants of $c_{\mathrm{X}}=44.3\pm 0.5%
\text{ }\mathrm{nm}/\mathrm{V}$ along $X$ axis and $c_{\mathrm{Y}}=32.9\pm
1.3\text{ }\mathrm{nm}/\mathrm{V}$ along $Y$ axis. In Table.\ref{T1}, we compare the uncertainties of calibration constants with different calibration methods. It indicates that the uncertainty of position measurement was about $1.0\%$ at the sensitivity level of $0.44$ $\mathrm{pm}/\sqrt{\mathrm{Hz}}$ (X-axis), which was much lower than that from other methods \cite{hebestreit2018,ricci2019mass}. Also, the operation pressure is $6$ orders of magnitude lower, which can be applied to enhance the sensitivity due to high environment noises isolation.

\begin{table*}[t]
\caption{Uncertainties of calibration constants with different calibration methods.}\centering%
\begin{ruledtabular}
\begin{tabular}{llll}
Calibration criterion & Primary uncertainty source & Operational pressure &
Relative uncertainty \\
\hline
Thermal stochastic force \cite{hebestreit2018} & Mass & $10$ mbar to $1$ atm & $15\%$ \\
Electrostatic force \cite{hebestreit2018}  & Mass & $<10$ mbar & $30\%$ \\
Stochastic and electrostatic force \cite{ricci2019mass}  & Electric-field strength & $\sim50$ mbar & $1.2\%$\footnotemark[1] \\
Potential nonlinearity [This work] & Potential geometry shape & $10^{-5}$ mbar\footnotemark[2] & $1.0\%$ \\
\end{tabular}%
\end{ruledtabular}
\\
\footnotetext[1]{The relative uncertainty is derived from its mass measurement uncertainty.}
\footnotetext[2]{The feedback amplitude locking is operational when $\delta\Gamma\gg\Gamma_{0}$, where $\delta\Gamma$ is the feedback introduced damping following $\delta \Gamma=\eta_{\text{m}}\Omega_{0}/(2\pi)$ \cite{zheng2019cooling} and $\Gamma_{0}$ is the air damping. It can be used in a pressure $<10^{-3}$ mbar }
\label{T1}
\end{table*}

Since we can get the calibrated trajectory of the trapped particle with high accuracy, it is
possible to extract more information about the levitated oscillator from its
thermal motion such as the calibration constants of other axes, mass and
density of the trapped nanoparticle.
For a nano-mechanical resonator sensor, the force measurement uncertainty
highly depends on the accuracy of the mass of the trapped
nanoparticle. Based on the calibration constant, we can obtain the mass of
the nanoparticle from the mean square displacement along one axis in a thermal equilibrium, which follows
\begin{equation}
m=\frac{k_{B}T}{\left\langle q_{\mathrm{X}%
}^{2}\right\rangle \Omega_{\mathrm{X0}}^{2}}\text{,}  \label{thermal}
\end{equation}%
where $k_{B}$ is the Boltzmann constant and $T$ is the
temperature of surrounding environment.

\begin{figure}[b]
\includegraphics[width=0.45\textwidth]{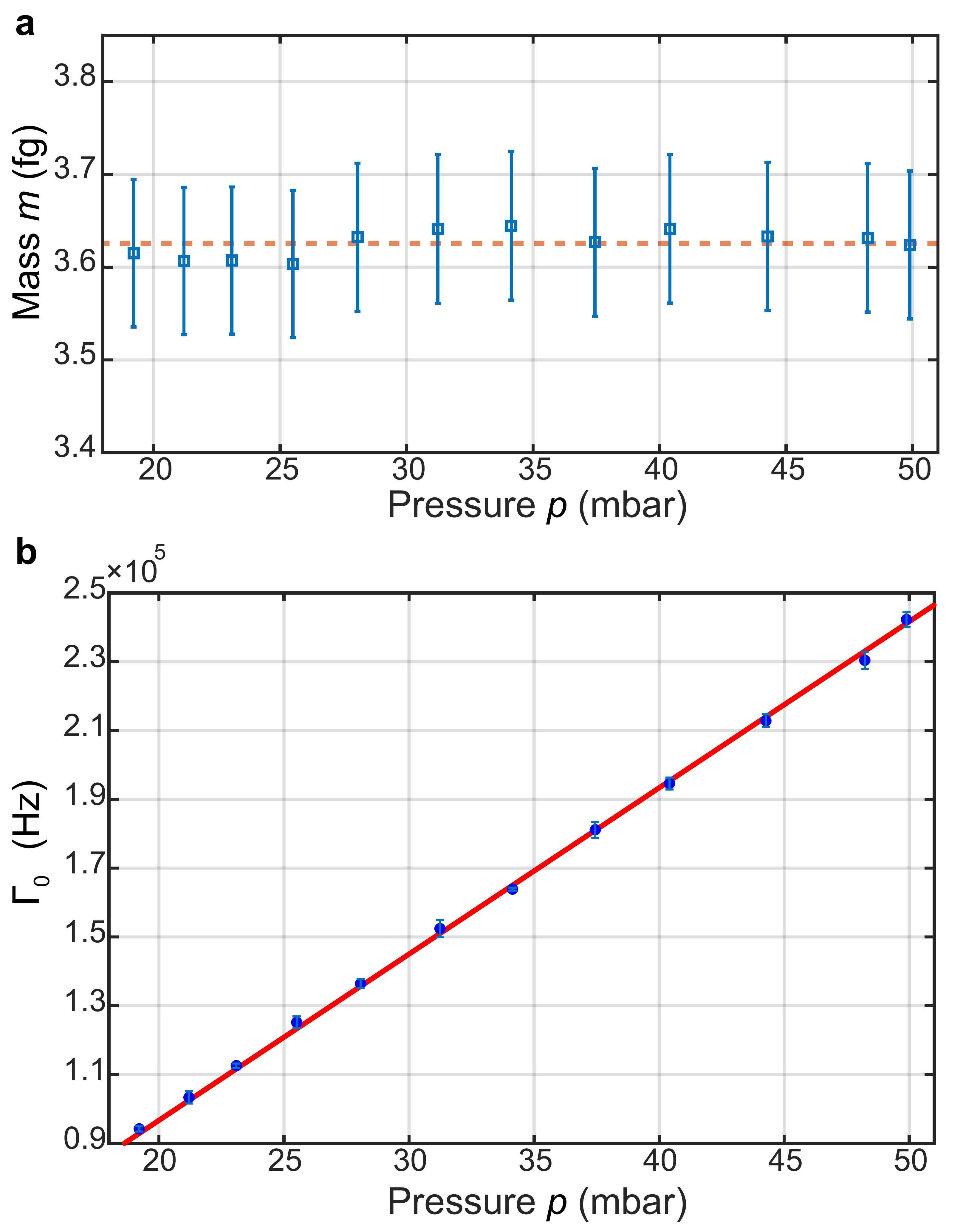}
\caption{(a) Measured mass under different air pressures. The error bars came from the systematic errors. The dashed line is the average of the measurement results. (b) Measured damping coefficients $\Gamma_{0}$, which were averaged from X, Y, and Z axes as a function of air pressure. The solid line is a linear fitting. The error bars came from the variance of $\Gamma_{0}$ between different axes.}
\label{fig:4}
\end{figure}

In the experiment, the thermal
motion properties, such as mean square signal voltage ($%
\left\langle V_{\mathrm{X}}^{2}\right\rangle =\left\langle q_{\mathrm{X}%
}^{2}\right\rangle /c_{\mathrm{X}}^{2}$), natural angular frequency ($\Omega_{\mathrm{%
X0}}^{2}$) and damping coefficient ($\Gamma_{0}$ along X, Y, and Z axes), were measured at a pressure between $20$ to $50$ \textrm{mbar}
by fitting the PSD. The pressure range we selected is to avoid the
measurement errors due to the nonlinear distortion of PSD, effective bath
temperature elevating under high vacuum and fitting error near overdamped regime. Before measurement, the vacuum chamber was exhausted
below $0.1$ \textrm{mbar} for a while and re-inflated to minimize the influence of particle
properties varying at low gas pressure like the evaporation of adsorbed
solvent. From the experiment measurement, we got the mass of the trapped
nanoparticle was $m=(3.63\pm 0.08)\times 10^{-18}$ \textrm{kg} with Eq. (\ref%
{thermal}) \cite{SI}.

Moreover, the density of nanoparticle is an important
parameter for its industrial and technological applications. Also the
density of a nanoparticle is related to its refractive index \cite{jackson1962classical} which is one of
the key parameters for optical levitation properties. However, it is difficult
to measure the density of an individual nanoparticle. And for the most popular St\"ober silica nanospheres, its density can be smaller than a bulk glass due to its porosity \cite{Parnell2016Porosity,Blakemore2019mass}. As for the density measurement, when the Knudsen number $\mathrm{Kn}\gg1$, where $\mathrm { Kn } = \overline { l } / R$, $\overline { l }$ is the mean free path of air molecules, and  $R$ is the radius of the nanosphere, the air damping coefficient of a nanosphere follows \cite{Epstein1924damping}
\begin{equation}
\Gamma _{0}=\frac{8}{3m}\left(1+\frac{a \pi}{8}\right) \sqrt{\frac{2 \pi m_{g}}{ k_B T}} p R^{2} \text{,}
\label{pauli}
\end{equation}%
where $p$ is the air pressure, $m_g$ is the mass of the air molecule, and $a$ is the momentum accommodation coefficient. With the measured damping coefficients in Fig. \ref{fig:4} and Eq. (\ref{pauli}), we can get that the radius of the trapped nanosphere was $%
R=75.4\pm 1.5\text{ }\mathrm{nm}$ and the density was $\rho =2.02\pm 0.10$ $%
\mathrm{g}/\mathrm{cm}^{3}$ \cite{SI}. This result agreed with the value provided by the manufacturer ($\rho_{\mathrm{Bangs}}\simeq2.0\text{ }\mathrm{g}/\mathrm{cm}^{3}$). As for a non-spherical nanoparticle, the
dependence of the damping rate on the pressure can be obtained by the direct simulation with Monte Carlo method \cite{Ahn2018GHz}.

In conclusion, we have introduced a nonlinear frequency shift based measurement of the position, mass and density of a nanoparticle with optical levitation in high vacuum.
We are able to control the amplitude of an optically levitated oscillator with tiny
deviation and make it possible to deploy a precise
nonlinear frequency shift measurement. Such a method does not
require the mass of nanoparticle or an assistance from an external force. The absolute precision is then mainly limited by the error from the tight focusing light field
estimation, which can be further measured with high precision in experiment \cite{bauer2014light,Neugebauer2015light,bautista2016light}. Moreover, an amplitude locked nano-oscillator can be regarded as a nearly ideal harmonic oscillator for classical and quantum investigation. It is possible to transform a position or velocity depended static interaction, such as Casimir force, electric field gradient, and Lorentz force, into a harmonic force, since an optically levitated nanoparticle sensor has an incredible sensitivity for resonant force measurement. Such an amplitude locked optically levitated nanoparticle can also be applied in the study of nonequilibrium physics and thermodynamics at the nanoscale \cite{jones2015tweezer,Gieseler2018therm}.

\section*{Methods}
\noindent{\textbf{Experimental setup.}} An objective ($\mathrm{NA}=0.9$%
) was mounted in the vacuum chamber to focus the $1064$ $\mathrm{nm}$
Gaussian beam laser ($\sim 200$ \textrm{mW}). Before the objective, the beam diameter was $4.5$ \textrm{mm}, which was larger than the back pupil
diameter ($3.6$ \textrm{mm}) of the objective to make full use of the
numerical aperture. An acousto-optic modulator (AOM) was mounted to modulate
the laser intensity based on the feedback control signal. The forward
scattering light was collected by an aspheric lens ($%
\mathrm{NA}=0.546$) and sent to three sets of balanced photodetectors to measure
the positions of the trapped nanoparticle along three motional degrees (set
as $X$, $Y$, $Z$ axis) of freedom. A silica nanosphere (nominal radius $82\pm10$
\textrm{nm}, Bangs labs Inc.) dispersed in ethanol was sent to the optical
trap with a nebulizer. The position voltage signals were recorded by a digitizer on computer and
simultaneously sent to a field programmable gate array (FPGA) board (Xilinx Spartan-6 XC6SLX16 Core Board) \cite{zheng2019cooling} to
generate the feedback modulation signal that can control the oscillation
amplitude along each axis.

\section*{Acknowledgment}
This work is supported by the Science Challenge Project (No. TZ2018003), the
National Natural Science Foundation of China (Nos. 91536219, 61522508, and
91850102), the Anhui Initiative in Quantum Information Technologies (No.
AHY130000).

\clearpage
\newpage
\leftline{\textbf{Supplementary Information}}
\section{Experimental setup}
\begin{figure*}[t]
	\includegraphics[width=0.8\textwidth]{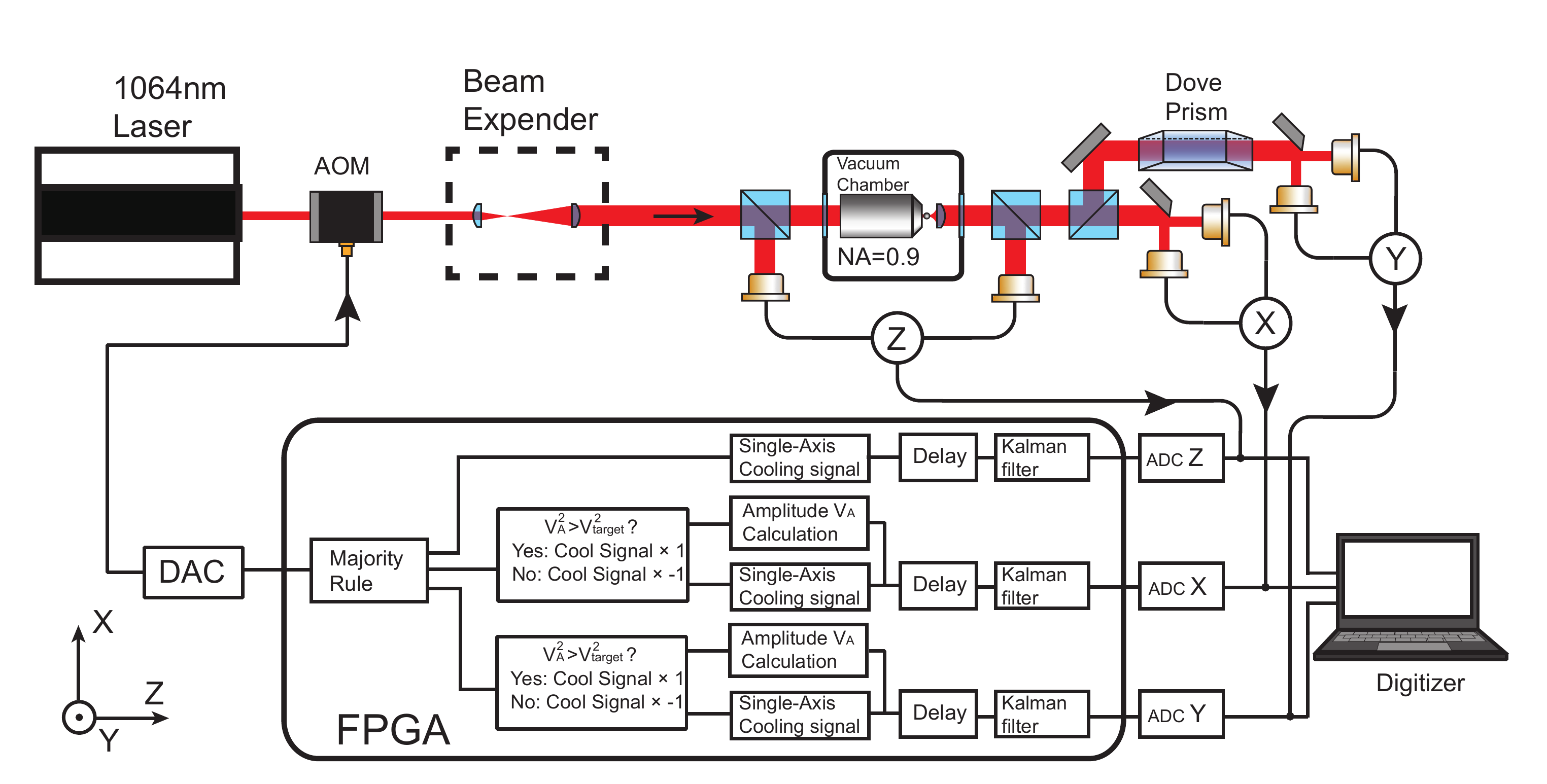}
	\caption{Experimental setup}
	\label{fig:S1}
\end{figure*}

As shown in Fig. \ref{fig:S1}, a $1064$ nm laser was used to trap the nanoparticle in vacuum chamber. Its intensity was modulated with an acousto-optical modulator (AOM). The laser beam was expanded to fulfill the back pupil of the objective (Nikon CFI LU Plan Fluor EPI 100X). The forward scattering light from the optical trap was collected by an aspheric lens (NA$=0.546$) and sent to three sets of balanced photodetectors to measure the positions of the trapped nanoparticle along three motional degrees (set as X, Y, Z axis) of freedom.

The position voltage signals were sent to a FPGA board to generate the feedback control signal. The digitized signal first passed a Kalman filter to suppress the out-band noise. And then, to eliminate the influence of feedback-loop time delay, two suitable delays were added to the voltage signal to generate a phase-matched signal $ u_{\mathrm{sin}} $ and a $\pi/2$ phase-shifted signal $ u_{\mathrm{cos}} $. $ u_{\mathrm{sin}} $ and $ u_{\mathrm{cos}} $ were sent to generate the feedback cooling signal with $ S_{cool}=\mathrm{sign}(u_{\mathrm{sin}}u_{\mathrm{cos}}) $ \cite{zheng2019cooling}. Simultaneously, the square of signal voltage amplitude $V_A^2$ was calculated with $ V_A^2=u^2_{\mathrm{sin}}+u^2_{\mathrm{cos}} $.  The amplitude was processed in square form because it is not convenient to do square root with FPGA. Then $V_A^2$ was compared with the target amplitude $V_{\mathrm{target}}^2$. If $V_A^2\geq V_{\mathrm{target}}^2$, the cooling signal $ S_{cool} $ would not be changed. Else if $V_A^2<V_{\mathrm{target}}^2$,  $ S_{cool} $ would be reversed and become a feedback heating signal. The feedback control signals along three axes were merged based on majority rule \cite{zheng2019cooling}. Finally, based on the modulation depth $ \eta_\mathrm{m} $, the merged three-dimensional control signal was processed and converted to an analog voltage signal to control the laser intensity with AOM.
\begin{figure*}[t]
	\includegraphics[width=0.8\textwidth]{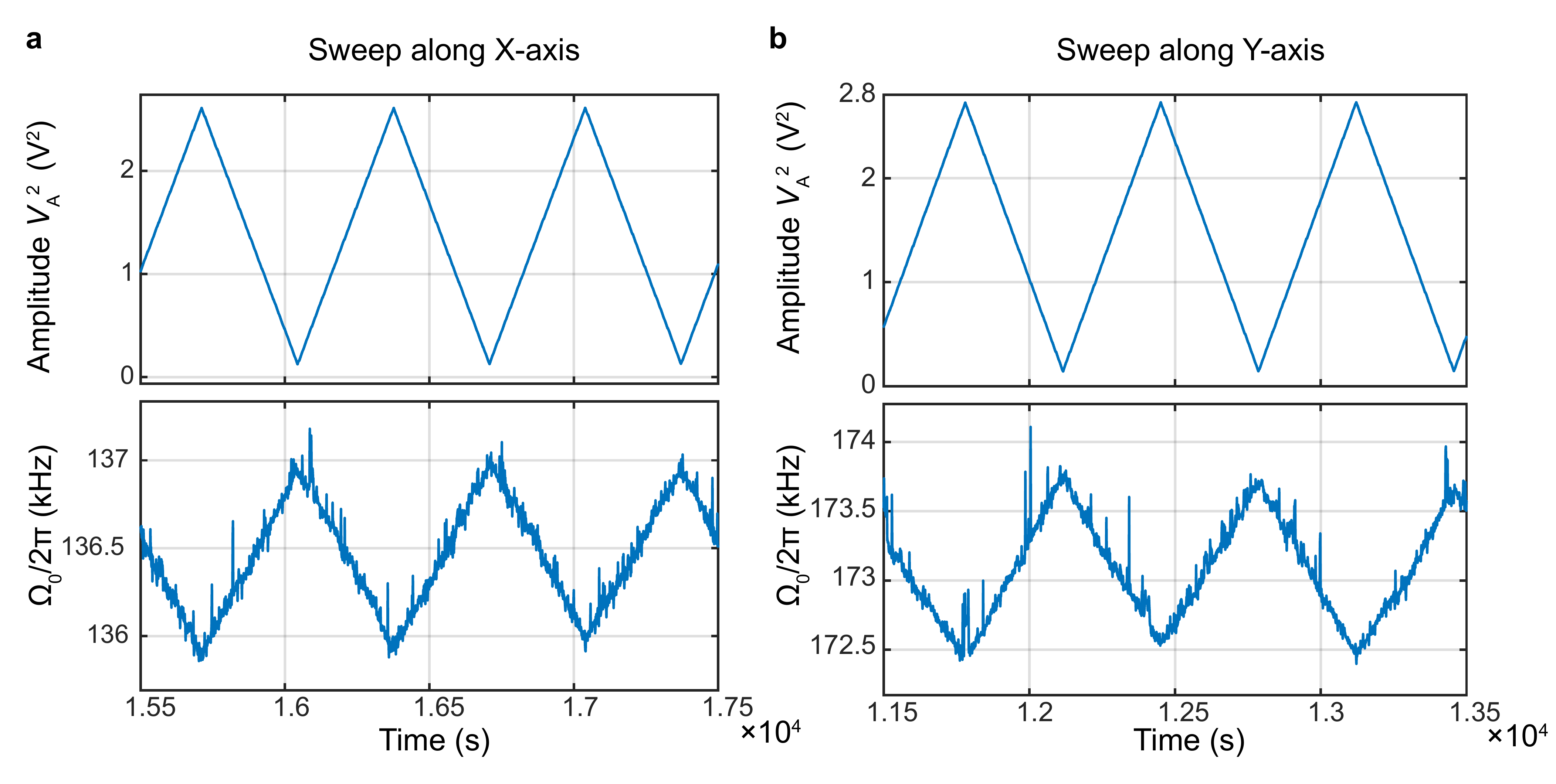}
	\caption{Part of the measured time trace of the swept voltage amplitude and natural frequency. (a) Sweep along X-axis. (b) Sweep along Y-axis. }
	\label{fig:S2}
\end{figure*}

During the calibration, to obtain the nonlinear coefficient $\alpha$, the square of voltage amplitude was swept between $0.16\text{ }\mathrm{V}^2$ to $2.56\text{ }\mathrm{V}^2$. This voltage sweeping was applied to X and Y axis motions as shown in Fig. \ref{fig:S2}. The frequency shift versus amplitude voltage was averaged from tens of sweeping cycles to reduce noises. Such calibration was not applied to Z-axis because the nonlinear property along Z-axis was different from X and Y axes due to the nonlinearity of the scattering force and tremendously affected by changes in equilibrium position. The equilibrium position was related to the radius and refractive index of the silica nanoparticle which were unable to obtain accurately.
\section{Light field estimation}
The Debye integral \cite{novotny2012principles} was utilized to make the simulation of the tightly focused light field around trapping position. The simulation condition we used was that the input laser beam has a Gaussian intensity distribution with a beam diameter of $4.5$ mm which was measured with a CCD camera beam profiler (newport LBP2). The back pupil diameter of the objective was $3.6$ mm. The laser wavelength was $1064$ nm. And the laser was linearly polarized along X axis. the numerical aperture (NA) of the objective was $0.9$. The equilibrium position along Z-axis was $ z_{eq}\simeq100 $ nm. We fitted the intensity distribution at $ z_{eq}=100 $ nm with Gaussian function along X and Y axes and with $I(z)=A/[1+(z+z_0)^2] $ along Z axis. The fitted $1/\mathrm{e}^{2}$ intensity radius along X and Y axes were $w_{\mathrm{X}}=696.6\text{ }\mathrm{nm}$ and $w_{\mathrm{Y}}=530.0\text{ }\mathrm{nm}$, and the Rayleigh range was $z_0=1080.0\text{ }\mathrm{nm}$.

According to the optical force with Rayleigh approximation and the light field estimation result, we were able to calculate the natural frequency ratio between each axes.
Based on the above light field estimation results, the natural frequency ratio between X, Y, and Z should be $1.31:1:2.93$, comparing with the experimental result $1.27:1:3.07$.

There were two possible reasons for the differences between the estimation and experimental results. First, the objective we used was designed for visible light. It introduced a slight deterioration of the focal length when it was used for near-infrared light. Second, the actual intensity distribution of the input laser was deviated from a perfect Gaussian distribution and there might be a slight misalign between the input laser and backpupil of the objective. It would introduce an error of the effective input laser diameter. The above reasons introduced simulation condition errors about the NA and the fill factor (input laser diameter/back pupil diameter). By tuning the simulation conditions, when the NA=$0.875$ and the input laser diameter was $4.2$ mm, the simulation result of the natural frequency ratio will be as same as the experimental result. Therefore, the simulated $1/\mathrm{e}^{2}$ intensity radius along X and Y axes were estimated to $w_{\mathrm{X}}=703\pm 7\text{ }\mathrm{nm}$ and $w_{\mathrm{Y}}=551\pm 22\text{ }\mathrm{nm}$. The errors demonstrated the difference between the experiment condition simulation and natural frequency ratio corrected simulation. Those errors can be further reduced in experiment \cite{bauer2014light,Neugebauer2015light,bautista2016light}.
\section{Error analysis of the mass, radius, and density measurement}
To measure the mass, radius, and density of the trapped nanosphere, a total of $12$ data points at a pressure between $20$ to $50$ mbar are recorded. The recorded information of each data point was derived by fitting the power spectral density (PSD) of position signals, including the mean square signal voltage, natural frequency, and damping coefficient along X, Y, and Z axes.

Because the calibration error along X axis was less than Y axis, the mean square signal voltage ($\left\langle V_{\mathrm{X}}^{2}\right\rangle$) and natural angular frequency ($\Omega_{\mathrm{X0}}$) along X axis was utilized to obtain the mass. Here,
\begin{equation}
m=\frac{k_{B}T}{c_\mathrm{X}^2\left\langle V_{\mathrm{X}%
	}^{2}\right\rangle \Omega_{\mathrm{X0}}^{2}}\text{,}
\label{thermal}
\end{equation}%
where $k_B$ is the Boltzmann constant, $T$ is the environment temperature, and $c_\mathrm{X}$ is the calibration factor along X. Based on the uncertainties shown in Table. \ref{density} and the propagation of uncertainty, we can get that $m=(3.63\pm 0.08)\times 10^{-18}$ \textrm{kg}.

\begin{table*}[t]
	\caption{\label{density}Uncertainties table.}\centering%
	\begin{ruledtabular}
		\begin{tabular}{lll}
			Quantity & Value $z_i$ & Error $\delta_{z_i}/z_i$ \\
			\hline
			$c_\mathrm{X}$&44.3 $\mathrm{nmV^{-1}}$&0.010\\
			$\Omega_{\mathrm{X0}}/(2\pi)$&135450 Hz&0.001(stat)\\
			$\left\langle V_{\mathrm{X}}^{2}\right\rangle$&0.7980 $\mathrm{V^2}$&0.005(stat)\\
			$\Gamma_{0}$&$(0.942\sim2.42)\times 10^{5}\text{ }\mathrm{s^{-1}}$&0.010\footnotemark[1]\\
			$p$&$(1.92\sim4.99)\times 10^{3}\text{ }\mathrm{Pa}$&0.002\footnotemark[2]\\
			$a$&0.9&0.111\\
			$T$&298 K&0.003\footnotemark[2]\\
			$m_g$&$4.8089\times 10^{-26}\text{ }\mathrm{kg}$&$-$\footnotemark[3]\\
			$k_B$&$1.3806\times 10^{-23}\text{ }\mathrm{JK^{-1}}$&$-$\footnotemark[3]\\
			\hline
			$m$&$3.63\times 10^{-18}\text{ }\mathrm{kg}$&0.022(sys), 0.004(stat) \\
			\hline
			$R$&$75.4\times 10^{-9}\text{ }\mathrm{m}$ &0.020(sys), 0.004(stat)\\
			\hline
			$\rho$&$2.02\text{ }\mathrm{g}/\mathrm{cm}^{3}$&0.052(sys), 0.013(stat)\\
		\end{tabular}%
	\end{ruledtabular}
	\\
	\footnotetext[1]{This error comes from the inconsistent damping coefficients between different axes due to the deviation from perfect sphere.}
	\footnotetext[2]{From manufacturer datasheets.}
	\footnotetext[3]{The error is less than 0.001.}

\end{table*}

As for the radius, the most often used formula to estimate the damping coefficient in optically levitation is \cite{Beresnev_Chernyak_Fomyagin_1990,Li2013}
\begin{equation}
\Gamma_{0}=\frac{6 \pi \eta R}{m} \frac{0.619}{0.619+K n}\left(1+c_{K}\right) \text{,}
\label{knudsen}
\end{equation}
where $R$ is the radius of the nanosphere, $c _ { K } = 0.31 \mathrm { Kn } / \left( 0.785 + 1.152 \mathrm { Kn } + \mathrm { Kn } ^ { 2 } \right)$, $\eta$ is the viscosity coefficient of air, and
$\mathrm { Kn } = \overline { l } / R$ is the Knudsen number.
$\overline { l } = k _ { B } T / \sqrt { 2 } \pi d ^ { 2 } p$ is the mean free path of air molecules, where $p$ is the air pressure and $d$ is the collision diameter of air molecules. And the correctness of Eq. (\ref{knudsen}) is under the condition that the momentum accommodation coefficient $a$ of the target sphere is equal to $1$. Some experiments have recommended that $a\sim0.9$ \cite{Ewart2007accommodation}.

So we turn to the Epstein's work that the damping coefficient in the free-molecule regime $(\mathrm {Kn}\gg1)$ is \cite{Epstein1924damping}
 \begin{equation}
 \Gamma _{0}=\frac{8}{3m}\left(1+\frac{\pi a}{8}\right) \sqrt{\frac{2 \pi m_{g}}{ k_B T}} p R^{2} \text{,}
 \label{Epstein}
 \end{equation}%
where
$m_g$ is the mass of the air molecule.
We can obtain the radius of the nanosphere with
\begin{equation}
{R}=\left(\frac{\Gamma_{0}}{p}\right)^{\frac{1}{2}}\left(\frac{8+\pi a}{3 m}\right)^{-\frac{1}{2}}\left(\frac{2 \pi m_{g}}{k_{B} T}\right)^{-\frac{1}{4}} \text{.}
\label{radius}
\end{equation}

Based on the uncertainties shown in Table. \ref{density} and the propagation of uncertainty, the radius of the trapped nanosphere was estimated to be $R=75.4\pm1.5$ nm.

As for the density, with Eq. (\ref{radius}), we have
\begin{equation}
\rho=\frac{3}{4 \pi}m^{-\frac{1}{2}}\left(\frac{\Gamma_{0}}{p}\right)^{-\frac{3}{2}} \left(\frac{8+\pi a}{3}\right)^{\frac{3}{2}}\left(\frac{2 \pi m_{g}}{k_{B} T}\right)^{\frac{3}{4}} \text{.}
\end{equation}
We can get the density of the trapped nanoparticle was $\rho=2.02\pm0.10\text{ }\mathrm{g}/\mathrm{cm}^{3}$.

\end{document}